\pdfoutput=1 

\documentclass[manuscript=letter]{achemso}
\setkeys{acs}{articletitle = true}
\usepackage{setspace}
\usepackage{amsmath}
\usepackage{amssymb}
\usepackage{mathtools}
\usepackage{soul}
\usepackage{multirow}
\usepackage{graphicx}	% Include figure files
\usepackage{dcolumn}		% Align table columns on decimal point
\usepackage{bm}	 	 	% bold math
\newcommand*{\citen}{}% generate error, if `\citen` is already in use
\DeclareRobustCommand*{\citen}[1]{%
  \begingroup
    \romannumeral-`\x % remove space at the beginning of \setcitestyle
    \setcitestyle{numbers}%
    \cite{#1}%
  \endgroup
}

\title{On the Nature of Trapped-Hole States in CdS Nanocrystals and the Mechanism of their Diffusion}
\author{R. Peyton Cline}
\author{James K. Utterback}
\author{Steven E. Strong}
\author{Gordana Dukovic}
\author{Joel D. Eaves}
\email{Joel.Eaves@colorado.edu}
\affiliation{Department of Chemistry and Biochemistry, University of Colorado Boulder, Boulder, Colorado 80309-0215, USA}

\begin{document}

\newpage
\onehalfspacing
\begin{abstract}
Recent transient absorption experiments on CdS nanorods suggest  that photoexcited holes rapidly trap to the surface of these particles and then undergo diffusion along the rod surface. In this paper, we present a semiperiodic DFT model for the CdS nanocrystal surface, analyze it, and comment on the nature of both the hole-trap states and the mechanism by which the holes diffuse. Hole states near the top of the valence band form an energetic near continuum with the bulk, and localize to the non-bonding sp$^3$ orbitals on surface sulfur atoms. After localization, the holes form nonadiabatic small polarons that move between the sulfur orbitals on the surface of the particle in a series of uncorrelated, incoherent, thermally-activated hops at room temperature. The surface-trapped holes are deeply in the weak-electronic coupling limit and, as a result, undergo slow diffusion.
\end{abstract}

% not working properly; leave out for now; not needed for arxiv
%\begin{tocentry} 
%\includegraphics{Peyton_TOC_fig} 
%CdS Surface Hole Hopping
%\end{tocentry}

In semiconductor colloidal nanocrystals, surface traps play a pivotal role in deciding the fates of photoexcited carriers. Despite their importance, the microscopic nature and character of the trapped states remains elusive. In cadmium-chalcogenide systems like CdS and CdSe, electron and hole traps form on the surface.\cite{brus1986, wuister2004, jasieniak2007, gomezcampos2012, peterson2014, krause2015, busby2015, gao2015, kilina2016, houtepen2017, wei2012} In most syntheses, capping ligands form  bonds with the cadmium atoms at the surface and turn electron trapping into a minor decay pathway.\cite{peterson2014, houtepen2017, rosenthal2007, knowles2010, sadhu2011, utterback2015} In contrast, chalcogen atoms at the surface often remain undercoordinated and participate in hole trapping.\cite{klimov1999, wu2015, gomezcampos2012, kilina2016, houtepen2017, wei2012, keene2014}
The valence band holes in these materials exhibit sub-picosecond lifetimes, the quantum yield for band gap photoluminescence is low, and there is a relatively strong emission from recombination between trapped holes and conduction band electrons.\cite{klimov1996, klimov1999, knowles2011, wu2015, peterson2014, krause2015, wu2012, mooney2013_2, guyotsionnest2005, bullen2006} The energies of the hole states relative to the bulk valence band and the extent of wavefunction localization within the trapped states of nanocrystals have been inferred from experimental spectra, though calculations for small clusters do exist.\cite{kilina2013, kilina2016, wei2012, houtepen2017} Recent work by Utterback {\it et al.}\cite{utterback2016} makes the picture richer, showing that trapped holes are not static, but are mobile at room temperature. That work showed that a one-dimensional diffusion-annihilation model could explain the long-lived power law decays in transient absorption spectra, and hypothesized that hole diffusion occurs through hopping, where hops are on the order of individual bond lengths. Deducing the mechanism for this hole motion and testing the hypothesis put forward in Ref. \citen{utterback2016} is a principal goal of this manuscript.

In this work, we use density functional theory (DFT) to describe the atomic structure of the nanocrystal surface and bulk states of CdS and the electronic structure of the trapped holes in the material.   We analyze the electronic structure of bulk and surface states within the system, and by parameterizing a tight-binding Hamiltonian from the DFT data, put forward a mechanism for the motion of trapped holes on the surface by computing hole hopping rates using Holstein small polaron/nonadiabatic Marcus theory.  We show that the rates are related to an anisotropic diffusion coefficient, and from the rate theory, propose a lower bound on the reorganization energy for hole hopping that is consistent with the experimental data reported in Ref. \citen{utterback2016}.

\begin{figure}[h!]
\centering
\includegraphics[scale=1]{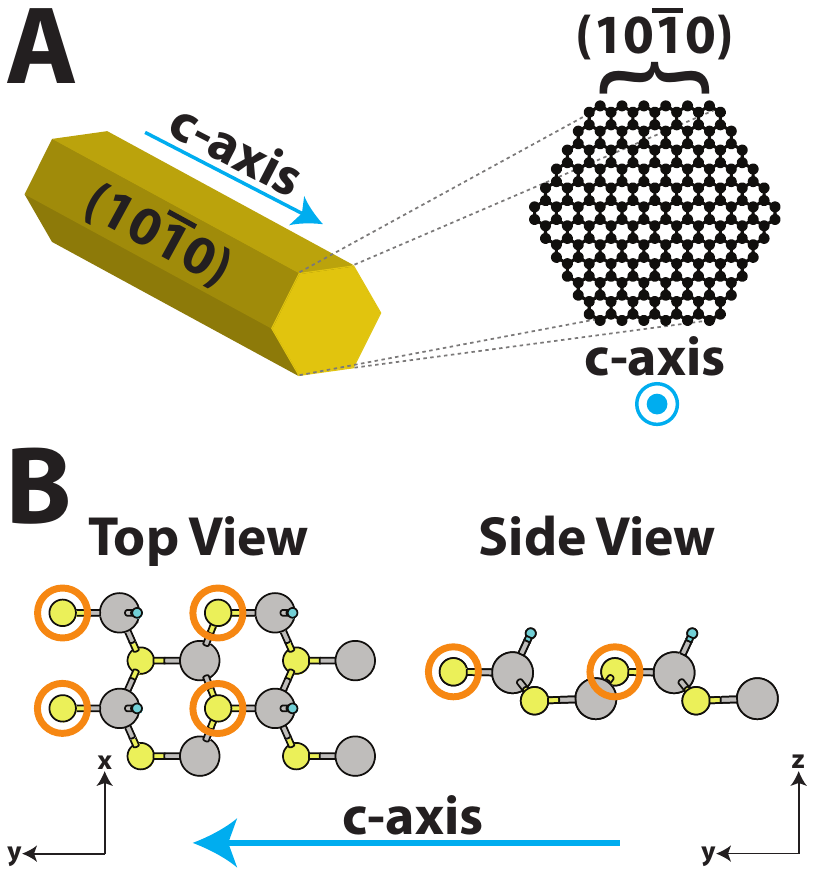}
\caption{\label{fig:schematic}(A) Illustration and cross section of a CdS nanorod. The cross section shows the 6 radial surfaces, which are all crystallographically the same and correspond to the $(10\overline{1}0)$ Miller-Bravais index. The crystal c-axis runs parallel to the long axis of the nanorod. (B) The surface atoms forming the top layer of the computationally relaxed supercell, showing 8 cadmium (grey) and 8 sulfur (yellow) from the top-down and side views. Pseudo-atoms, used as a surrogate for passivating ligands, appear in blue.  Orange circles outline the surface sulfur atoms. The remaining atoms forming the supercell are not shown (see SI, Fig. S2, for the full structure). The two small axes in black denote the orthorhombic coordinates mentioned in the text.}
\end{figure}

As Fig. \ref{fig:schematic}A illustrates, CdS nanorods exhibit 6 identical faces, with all facets characterized by the $(10\overline{1}0)$ Miller-Bravais index.\cite{rantala1996,rosenthal2007,widmercooper2014}
On this surface, both cadmium and sulfur atoms make three bonds to the crystal lattice, leaving one orbital on the atoms exposed to the surface to be either passivated by a ligand or left as a dangling bond. 

To construct a computational model for the CdS bulk and surface where ligands passivate cadmium but not sulfur on the surface\cite{utterback2016}, we begin by optimizing the nuclear positions of a $2\times 2\times 2$ wurtzite supercell in the bulk phase by minimizing the energy, changing both atomic positions and cell dimensions, at constant stress.  To build the surface ``slab'' supercell from the bulk wurtzite cell, we use the fact that a wurtzite crystal cleaved along the $(10\overline{1}0)$ direction has an orthorhombic symmetry,\cite{rantala1996,sun2013} with the sulfur atoms forming a rectangular lattice along in the orthorhombic x-y plane for CdS (Fig. \ref{fig:schematic}B). We choose coordinates so that the orthorhombic x-y plane coincides exactly with the wurtzite crystal b-c plane.   From the $2\times 2\times 2$ bulk supercell, we form the surface slab by replicating the $2\times 2\times 2$ bulk cell along the orthorhombic z-axis. It is convenient to define a layer as comprising 8 atoms of cadmium and 8 atoms of sulfur (Fig. \ref{fig:schematic}B) lying approximately in the orthorhombic x-y plane.  The results reported here use an 8-layer slab supercell, $2\times 2\times 8$ unit cells in size, though the SI provides extensive convergence tests as a function of supercell dimensions. This supercell generates two equivalent, but independent, $(10\overline{1}0)$ surfaces separated by a bulk phase of CdS sandwiched between the surfaces in a ``middle-out'' geometry. Finally, placing a vacuum layer 30 {\AA} thick normal to the $(10\overline{1}0)$ surface allows plane-wave basis functions and periodic boundary conditions to be applied in all directions, but ensures that periodic images in the z-direction do not interact with one another.

We use pseudo-hydrogen potentials\cite{deng2012, zhang2016} bonded to all surface cadmium sites to saturate the dangling bonds on cadmium atoms and mimic the effect of ligand passivation. To model the appropriate experimental systems, we leave surface sulfur atoms unpassivated. We then find the equilibrium nuclear positions for the slab by minimizing the energy, changing atomic positions at fixed supercell dimensions. The positions of the atoms in the middle 2 layers of the slab remain fixed during surface relaxation to simulate a rigid bulk region, though after relaxation the constraint forces on these atoms are zero to within accepted tolerance (23 meV/{\AA}). This implies that the middle 2 layers appear approximately as bulk to the other atoms, whose positions we allow to change until their atomic forces have magnitudes less than 10 meV/{\AA}. 

\begin{figure}[h!]
\centering
\includegraphics[scale=1]{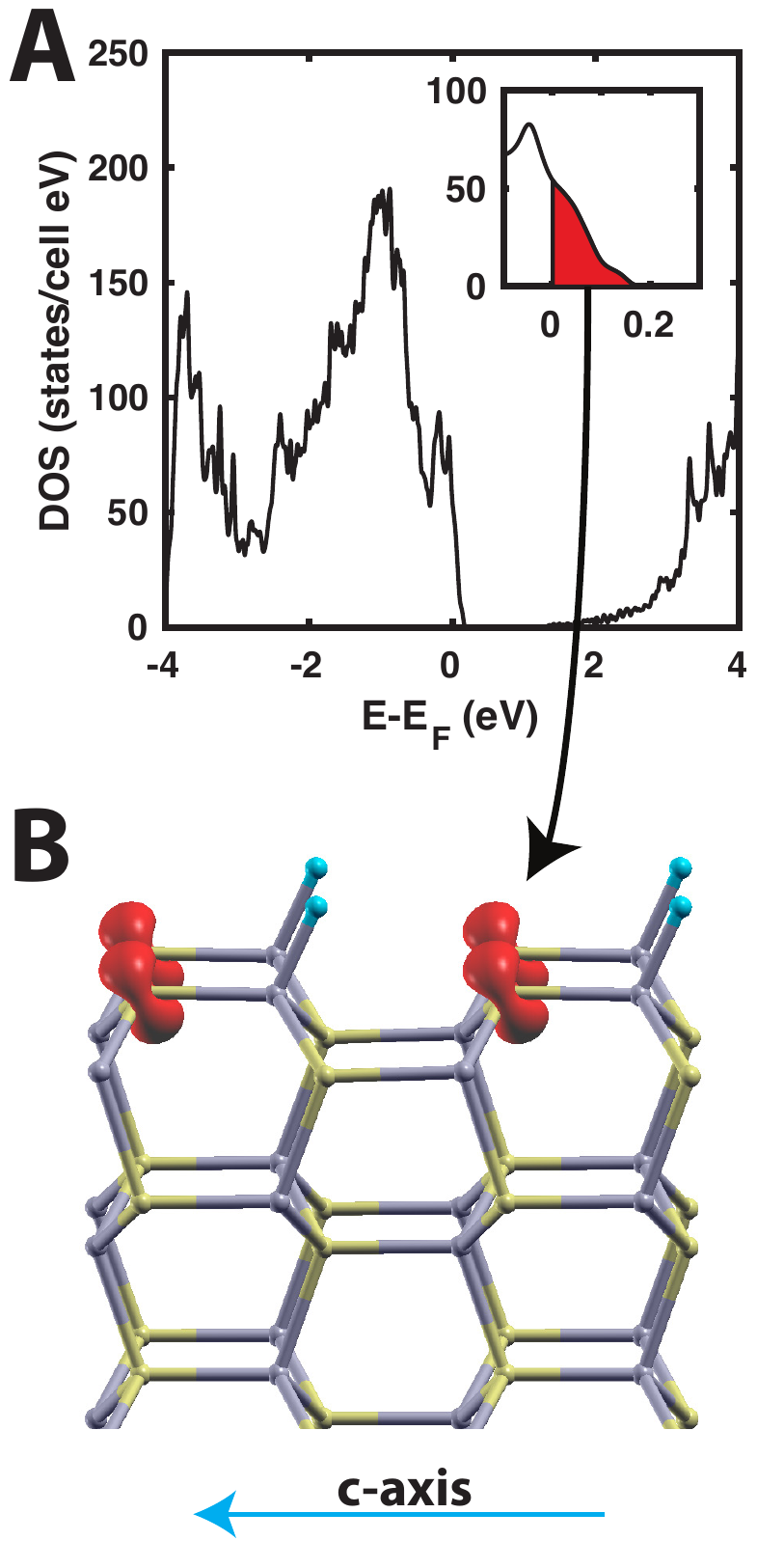}
\caption{\label{fig:DOS_PARCHG}(A) Total density of states (DOS) for the $2\times 2\times 8$ slab supercell, showing states within the range $E_F \pm 4$ eV, where $E_F$ is the Fermi energy. The inset focuses on the states above $E_F$ that are unoccupied at zero Kelvin and are therefore hole states.  (B) The hole-state density (red) corresponding to the energy range in the red part of the inset above shows holes are localized to the sulfur atoms on the surface. Only the top 2.5 layers of the supercell are shown. Focusing on the thermally accessible hole states within the energy range within $k_B T=25$ meV above or below $E_F$ exhibits a similar density. }
\end{figure}

Fig. \ref{fig:DOS_PARCHG}A shows the total density of states (DOS) of the slab around the Fermi energy, $E_F$.  
In these  calculations, $E_F$ identifies the energy at which the bands become half filled, similar to $E_F$ in a metal at finite temperatures.  At energies below $E_F$, the electron population dominates, while at energies above $E_F$, the hole population dominates. The inset to the DOS in Fig. \ref{fig:DOS_PARCHG}A  shows the region above $E_F$ but below the band gap, representing the lowest-energy hole states. Fig. \ref{fig:DOS_PARCHG}B shows the orbital density corresponding to the lowest energy hole states, which are clearly localized to the surface. This density looks like it could be formed from a linear combination of sp$^3$ hybridized orbitals, centered at the sulfur atoms located on the surface. Indeed, as we will show below, this observation provides the basis for a quantitative tight-binding model for the surface hole states. Constraining the energy window to lie within $k_B T = 25$ meV above or below $E_F$, to focus on the thermally accessible hole states, does not change the picture for the orbitals in Fig. \ref{fig:DOS_PARCHG}B qualitatively.  

Other electronic structure calculations of small CdS and CdSe clusters have reported that hole states are localized to the sulfur atoms on the surface, in agreement with this work.\cite{wei2012,houtepen2017,kilina2013,kilina2016} Unlike those calculations, however, our hole states are not well separated from the bulk DOS, but rather form a continuum with the highest fully-occupied (by electrons) states.  This result is consistent with the experimental work of Mooney {\it et al.},\cite{mooney2013_2} which finds that bulk and surface states can be separated by merely tens of meV.

The continuous density of states in the vicinity of the highest occupied valence bands, shown as the inset in Fig. \ref{fig:DOS_PARCHG}A, implies that holes prepared in the bulk rapidly move to the surface, lowering their energy by accessing a tier of intermediate states that are separated by infinitesimally small amounts of energy. This result supports experimental reports on CdS nanocrystals, which find that hole trapping to the surface occurs on the timescale of picoseconds \cite{wu2012,klimov1996,klimov1999,wu2015,utterback2016}.  

At first blush, the data in Fig. \ref{fig:DOS_PARCHG} seem to imply that the hole states are delocalized across sulfur atoms on the surface. But one must keep in mind the limitations of adiabatic electronic structure theory, which gives the solutions to the electronic structure problem in the Born-Oppenheimer approximation at zero absolute temperature. Similar to constructing molecular orbitals from linear combinations of atomic orbitals, if two orbitals are coupled, even weakly, DFT will yield adiabatic states that are delocalized superpositions.\cite{sakurai} But if the electronic coupling is weak relative to the coupling between the electrons and the environment, this delocalization will not survive decoherence from thermal fluctuations at finite temperature.\cite{mahan,weiss} Thermal decoherence in the electronic space will produce localized states that, while not eigenstates of the electronic Hamiltonian, more closely resemble the eigenstates of the system plus the environment.

\begin{figure}[h!]
\centering
\includegraphics[scale=1]{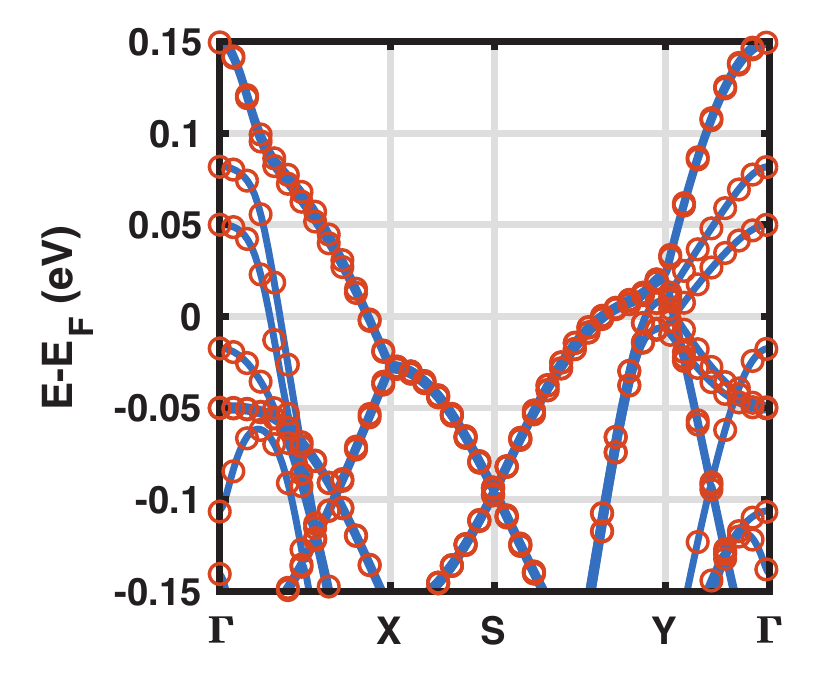}
\caption{\label{fig:BS}Band structure of the surface slab near the Fermi energy, $E_F$.  The original bands generated directly from DFT (blue lines) and the band structure from the eigenvalues of ${\cal H}_{TB}$ (orange circles) are overlaid to emphasize the accuracy of the Wannier localization procedure and the choice of sulfur sp$^3$ orbitals as the basis set for ${\cal H}_{TB}$. Symmetry points match the reciprocal directions of the x-y plane in the orthorhombic cell. The chosen energies include the energy range specified in the inset to the DOS in Fig. \ref{fig:DOS_PARCHG}A, which is dominated by surface sulfur states. }
\end{figure}

The calculations presented here do not explicitly include solvent, ligand, or electron-phonon coupling. While it is computationally infeasible to include them directly into the electronic structure calculations, we can include their effects by constructing a model Hamiltonian.  We begin by quantizing the electron density in terms of the sp$^3$ orbitals on all sulfur atoms. This leads to a tight-binding Hamiltonian, ${\cal H}_{TB}$, whose diagonal elements are the on the on-site energies, $\{ \epsilon_{\alpha} \}$, and the off-diagonal elements are the tunneling matrix elements, $\{ t_{\alpha,\beta} \}$, where the subscripts denote the sp$^3$ orbitals of the sulfur atoms. We use Wannier localization to generate the orbitals and compute the matrix elements. Fig. \ref{fig:BS} shows that, indeed, diagonalizing ${\cal H}_{TB}$ and forming the two-dimensional band structure for the directions conjugate to the x-y orthorhombic axes yields a band structure that is indistinguishable from the adiabatic solution to the electronic structure near the valence band maximum, where the surface hole states dominate. 

At zero temperature and in the absence of electron-phonon coupling, the adiabatic solution corresponding to the eigenvalues and eigenstates of ${\cal H}_{TB}$ would be an accurate solution to the electronic problem. But the energies of the relevant tunneling matrix elements for hole transport, described in more detail below and reported in detail in the SI (Tables S1-S2), are on the order of or are smaller than $k_B T$ at room temperature, implying that thermal fluctuations of the nuclei do indeed destroy coherence between orbitals and collapse the band structure in Fig. \ref{fig:BS}.\cite{alexandrov,mahan}

\begin{figure}[h!]
\centering
\includegraphics[scale=1]{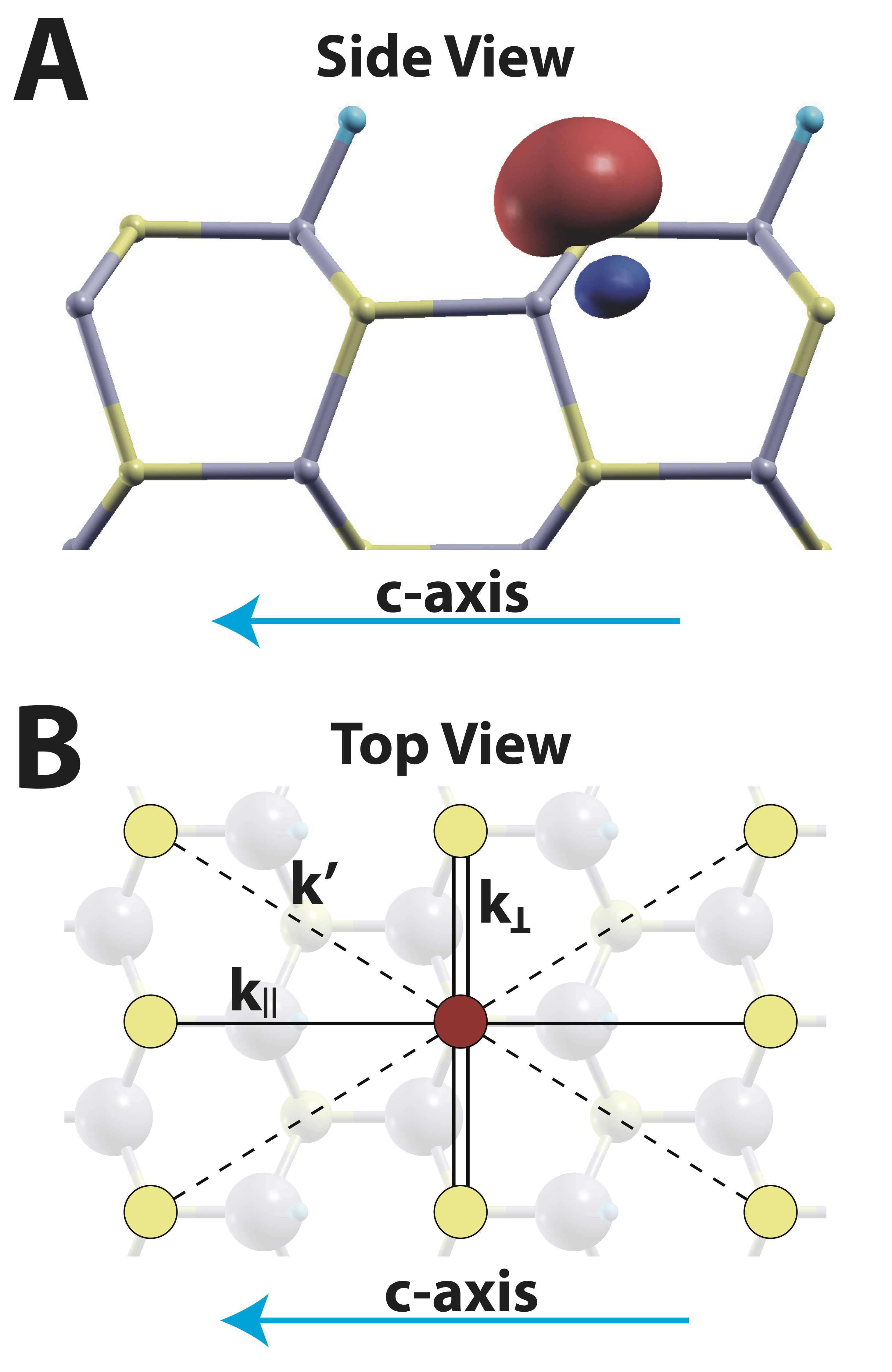}
\caption{\label{fig:rate_diabat}The relevant states used to construct the effective tight-binding Hamiltonian for surface holes. (A) Side view of an sp$^3$ orbital centered on a surface sulfur atom of the slab supercell. (B) A top-down view of the surface layer.  A hole that exists at the center position (red) can hop to any of the 8 labeled sites (yellow).  The rates $k_{\parallel}$ (single solid lines) and $k'$ (dashed lines) contribute to the diffusion constant for hole hopping down the crystal c-axis. }
\end{figure}

To construct a model Hamiltonian that captures these effects at the level of a linear response theory, we project out the surface sulfur atoms from ${\cal H}_{TB}$, giving an effective two-dimensional tight-binding Hamiltonian for the holes on the surface, ${\cal H}_S$. 
The system-bath Hamiltonian, ${\cal H}_{S-B}$, which couples the surface electronic degrees of freedom to the phonons, ligands, and solvent, is linear in the electron density, ${\cal H}_{S-B} = -\int d{\bm r} \rho({\bm r}) \Phi(\bm r)$, where $\Phi({\bm r})$ is the electrostatic potential of the nuclei.\cite{alexandrov,mahan} The electron density, quantized in the mutually-orthogonal Wannier sp$^3$ orbitals, $\{\psi_n({\bm r})\}$, is $\rho({\bm r}) = \sum_n c_n^\dag c_n |\psi_n({\bm r})|^2$. Assuming that all nuclei undergo harmonic vibrations about their equilibrium positions and expanding ${\Phi({\bm r})}$ to linear order in all bath modes leads to the Holstein small-polaron Hamiltonian, where the nuclear motions comprise a heat bath,
\begin{equation}
{\cal H} = \sum_{\langle \langle m,n \rangle \rangle} \epsilon_m c_m^\dag c_m + t_{m,n}c_m^\dag c_n + t_{n,m} c_n^\dag c_m + \sum_{n,\nu} \lambda^n_\nu(a_\nu^\dag + a_\nu)c_n^\dag c_n + \sum_\nu \hbar \omega_\nu a_\nu^\dag a_\nu.
\end{equation}
Here, $c_m^\dag$ ($c_m$) is the fermionic creation (annihilation) operator that puts (removes) a hole in orbital $m$, $a^\dag_\nu$ ($a_\nu$) is the bosonic creation (annihilation) operator for bath mode $\nu$, $\lambda_\nu^n$ is the coupling matrix element between mode $\nu$ and electronic orbital $n$, and $\hbar \omega_\nu$ is the energy of bath mode $\nu$. The sum over $m$ and $n$ goes over non-bonding orbitals of nearest- and next-nearest-neighbors on the surface. Fig. \ref{fig:rate_diabat}A shows one of these orbitals.

The relevant tunneling matrix elements for hole transfer down the crystal c-axis are $t_\parallel$ and $t'$, corresponding to the tunneling matrix elements between nearest- and next-nearest-neighbors along the rod axis, respectively (Fig. \ref{fig:rate_diabat}B). Our calculations for $t_\parallel$ and $t'$ yield  $|t_{\parallel}| \approx |t'| \approx 3$ meV. These small values for the tunneling matrix elements, compared to both room-temperature $k_B T$ and to characteristic values for $\lambda_\nu^n$,\cite{alexandrov,mahan} place this system squarely in the weak electronic coupling limit, where bath fluctuations destroy the band structure and localize holes to orbitals on individual sulfur atoms.  In this limit, the hole populations on the surface sites obey the quantum Master equation,
\begin{equation}
\frac{d P_{m}(t)}{dt} = \sum_n k_{n\rightarrow m}P_{n}(t) - \sum_{n}k_{m\rightarrow n} P_{m}(t) \label{eq:master} ,
\end{equation} 
with rates $k_{n \rightarrow m}$ given by the Golden Rule, $k_{n \rightarrow m}\sim |t_{m,n}|^2$. In the high temperature limit, which is the simplest approximation, bath energies are much less than $k_B T$, and the detailed spectrum of the heat bath is irrelevant. One then recovers the Marcus expression\cite{tachiya1993} for nonadiabatic hole transfer between degenerate states, 
\begin{equation}
k_{n\rightarrow m} =\frac{|t_{m,n}|^2}{\hbar} \sqrt{ \frac{\pi}{\lambda k_{B} T } } \text{exp}\Bigg\{ -\frac{\lambda}{4 k_{B} T}\Bigg\}\label{eq:marcus} ,
\end{equation}
where $\lambda = \sum_\nu\frac{(\lambda_\nu^n - \lambda_\nu^m)^2}{\hbar \omega_\nu}$ is the reorganization energy.  As implied earlier by our notation, $t_{\parallel}$ and $t'$ correspond to the rates $k_{\parallel}$ and $k'$, respectively (Fig. \ref{fig:rate_diabat}B).
Finally, an asymptotic expansion of Eq. \ref{eq:master}, presented in the SI (Sec. S.1), gives a Fokker-Planck approximation to the master equation, which is an anisotropic diffusion equation. The diffusion constant for hole hopping along the axis of the nanorod, $D_{\parallel}$, is 
\begin{equation}
	D_\parallel = c^2 ( k_{\parallel} + 2k' ),\label{eq:Deff}
\end{equation}
where $c = 6.8417$ {\AA} is the surface sulfur-to-sulfur distance parallel to the long axis of the nanorod. %The diffusion constant for hole hopping around the waist of the rod, $D_\perp$, does not contribute to electron-hole recombination along the axis of the nanorods.\cite{utterback2016}  
Note that $k'$, the hopping rate along the diagonal line connecting next-nearest-neighbors (Fig. \ref{fig:rate_diabat}B), makes a significant contribution to $D_\parallel$. The factor of two in Eq. \ref{eq:Deff} reflects the fact that there are two next-nearest-neighbors for each nearest-neighbor on the rectangular lattice, in the direction of the crystal c-axis. 

Calculating $\lambda$ is not an objective of this manuscript, nor is it computationally feasible in a system of this size.  Without a value for $\lambda$, we cannot compute the diffusion constant $D_{\parallel}$, but we can use our computed hopping matrix elements, $|t_{\parallel}|\approx |t'|\approx 3$ meV, in conjunction with experimental data to bound the value of $\lambda$.  Utterback {\it et al.}\cite{utterback2016} measured an upper bound for $D_{\parallel}$ to be $\sim$10$^{-7}$ cm$^2$s$^{-1}$. From this bound on $D_{\parallel}$ and our calculated values of $t_{\parallel}$ and $t'$, we use Eq. \ref{eq:marcus}-\ref{eq:Deff} find a lower bound for $\lambda$ of 1 eV.  This value of $\lambda$ is large for electron transfer reactions in solution, but not uncommon for polaron hopping transport in polar semiconductors, where lattice phonons couple strongly to the electronic degrees of freedom.\cite{deskins2007, mooney2013_2} Similarly, the high-temperature approximation to the system-bath coupling, while appropriate for a wide class of electron transfer reactions in solution, may need to be revised as more information about the specific coupling to lattice phonons and to nuclear distortions of the ligands, whose energies are not necessarily small relative to $k_B T$\cite{mooney2013_2}, becomes available. The tunneling matrix element for hopping perpendicular to the crystal c-axis, $|t_{\perp}|\approx 80$ meV (Fig. \ref{fig:rate_diabat}B), is about an order of magnitude larger than $t_{\parallel}$. Because it is still much smaller than $\lambda$, so the holes still remain small polarons localized to atomic sites, not delocalized around the waist of the rod.  

In summary, we have presented an atomistic computational model to capture the electronic structure of the bulk and surface states of CdS nanorods. The adiabatic solution to the electronic structure of this surface slab clearly shows hole states, near the Fermi energy, that are localized to the undercoordinated sulfur atoms on the surface.  Our results show that the hole states of a CdS surface slab form a near continuum with the bulk states of electrons, a result that differs from previous theoretical work on small quantum dots,\cite{wei2012, houtepen2017, kilina2013, kilina2016} but is consistent with recent experimental work of Mooney {\it et al.}\cite{mooney2013_2}  The continuous density of states between holes and electrons would also rationalize a wealth of experimental data that finds holes prepared in the bulk relax quickly to the surface.\cite{wu2012, klimov1996, klimov1999, wu2015, utterback2016} The densities of states for small clusters of CdS, which sometimes show trapped states that are well-separated from the valence band,\cite{wei2012, houtepen2017, kilina2013, kilina2016} differ from the density of states for the nanocrystal model presented here.

Motivated by the adiabatic electronic structure results, we parameterize a model tight-binding Hamiltonian from the sp$^3$ orbitals centered on sulfur atoms. The calculated tunneling matrix elements between relevant sulfur orbitals on the surface are small relative to the estimated reorganization energy, so that thermal fluctuations and nuclear motions localize holes into small polarons, trapped in non-bonding sulfur orbitals on the surface. While we cannot rule out hole hopping through a super-exchange mechanism, mediated through virtual bulk or ligand states, it is notable that this mechanism need not be invoked. Because CdS nanocrystals in conventional syntheses are so deep in the weak electronic coupling regime, hole transport in these particles is slow at room temperature. Our results give direct support for the picture presented by Utterback {\it et al.}\cite{utterback2016}, who hypothesized that holes diffuse by hopping on length scales comparable to interatomic spacings. 

\section{Computational Methods}
%
%vasp
%
The DFT calculations employed the  Vienna Ab initio Simulation Package\cite{kresse1993,kresse1994,kresse1996_2,kresse1996_1}, a plane-wave electronic structure code. We used the projector augmented wave method with the Perdew-Burke-Erzhenhoff generalized gradient approximation for the exchange-correlation functional,\cite{blochl1994,kresse1999} which for CdS strikes a reasonable balance between computational feasibility and accuracy.  The cadmium 5s$^2$4d$^{10}$ and sulfur 3s$^2$3p$^4$ electrons were treated explicitly in all calculations. 
%
%wannier
%
To analyze the results of the electronic structure calculation in the framework of nonadiabatic polaron theory, we employed a localization procedure using Wannier90\cite{mostofi2008}. Wannier90 encompasses a method for obtaining maximally-localized Wannier functions (MLWFs) from the adiabatic Bloch states, and it computes these MLWFs according to the method of Marzari and Vanderbilt.\cite{marzari1997,souza2001}  
%
%xcrysden
%
We used the XCrySDen\cite{xcrysden} software to visualize all structures, wavefunction densities, and Wannier functions. 

\begin{acknowledgement}
%
%J.D.E. CAREER AWARD:
%
This material is based upon work supported by the National Science Foundation under Grant No. CHE-1455365.
%
%SUMMIT:
%
This work utilized the RMACC Summit supercomputer, which is supported by the National Science Foundation (awards ACI-1532235 and ACI-1532236), the University of Colorado Boulder, and Colorado State University. The Summit supercomputer is a joint effort of the University of Colorado Boulder and Colorado State University.
%
%NSF FELLOWSHIPS:
%
J.K.U. and S.E.S. acknowledge support from National Science Foundation Graduate Research Fellowships under Grant No. DGE 1144083.   
%
%G.D. SUPPORT:
%
G.D. acknowledges support from Air Force Office of Scientific Research under AFOSR award No. FA9550-12-1-0137.
\end{acknowledgement}

\section{Supporting Information}
Derivation of the diffusion constants from the hopping rates; crystal structure parameters; bulk and surface convergence details; surface construction details; Wannier localization procedure details; additional figures. 

%\begin{suppinfo}
%Derivation of the diffusion constants from the hopping rates; crystal structure parameters; bulk and surface convergence details; surface construction details; Wannier localization procedure details; additional figures.
%\end{suppinfo}

\bibliography{researchBib}

\end{document}